# Creating Temporally Correlated High-Resolution Profiles of Load Injection Using Constrained Generative Adversarial Networks

Hritik Gopal Shah
Resiliency and Reliability Department
Eversource Energy, Hartford, CT, USA
hshah59@asu.edu

Behrouz Azimian, Anamitra Pal
School of Electrical, Computer, and Energy Engineering
Arizona State University, Tempe, AZ, USA
bazimian@asu.edu; Anamitra.Pal@asu.edu

*Abstract*—Traditional smart meters, which measure energy usage every 15 minutes or more and report it at least a few hours later, lack the granularity needed for real-time decision-making. To address this practical problem, we introduce a new method using generative adversarial networks (GAN) that enforces temporal consistency on its high-resolution outputs via hard inequality constraints using convex optimization. A unique feature of our GAN model is that it is trained solely on slow timescale aggregated historical energy data obtained from smart meters. The results demonstrate that the model can successfully create minute-by-minute *temporally correlated profiles of power usage* from 15-minute interval *average power consumption information*. This innovative approach, emphasizing inter-neuron constraints, offers a promising avenue for improved high-speed state estimation in distribution systems and enhances the applicability of data-driven solutions for monitoring and subsequently controlling such systems.

## I. Introduction and Motivation

Distribution systems were once considered passive components of the grid due to the absence of generation within them. However, with the increasing installation of distributed energy resources (DERs), distribution systems are transforming from passive elements to active market-ready entities. At the same time, the inherent variability of DERs poses a significant challenge to the reliable operation of active distribution systems [1], necessitating an evolution of their monitoring infrastructure. In this regard, the introduction of advanced metering infrastructure (AMI) in the form of *smart meters* has ushered in a new era, furnishing additional measurement sources that significantly amplify the observability of distribution systems. At the same time, it must be pointed out that AMI data is primarily used to perform energy-related tasks, such as monthly billing calculations. As such, their records capture *the average power consumption* over 15 or 60-minute intervals [2]. However, this aggregated reading is unable to capture the fast variations in *instantaneous power usage* values. More importantly, smart meters send data after a latency period that can range from *a few hours to a day* [3]. This means that AMI data are not available for real-time decision-making, which limits their use for important applications such as high-speed distribution system state estimation (DSSE) and voltage control [4]. Note that the terms averaged power and aggregated power used in this paper refer to the same thing, namely, power reported by smart meters/AMI over a given time interval.

Recently, it has been demonstrated that *learning-based methods* can enhance utilization of all available measurements in distribution systems by segregating the use of different time resolution data between offline training and real-time operation [3]-[6]. Historical AMI data can be employed in the offline training stage, thus overcoming the aforementioned limitations of smart meter measurements, while a limited set of high-speed data coming from sensors such as supervisory control and data acquisition (SCADA), phasor measurement unit (PMU), or micro-PMU, located at the feeder-head, can be used exclusively during online operation to bypass the need to fully observe the system in real-time using the high-speed sensors alone. In this way, the learning-based methods can provide fast and accurate state estimates for real-time unobservable distribution systems [7], while out-performing conventional methods that rely on pseudo-measurements achieved from load forecasting techniques [8].

A vital component of these learning-based methods that rely on fast timescale measurements during their online operation, is their *training database*. This is because the scenarios present in the database must effectively represent the load profile variations occurring at those fast timescales. Unfortunately, for creating such a database, one needs load injection profiles at a higher resolution than what AMI can provide. To address this practical problem, *data-driven generative models* can be employed to create fast timescale measurements from slow timescale measurements, similar to the concept of *super-resolution* [9]. Super-resolution, a well-known theory in image processing, involves generating high-resolution images from low-resolution images. In power system applications, Liu et al. introduced super-resolution perception for processing AMI data [10]. However, they

This work was supported in part by the National Science Foundation (NSF) under grant ECCS-2145063.

oversmoothed the reconstructed high-resolution data resulting in unrealistic profiles.

There is a growing body of literature that is employing generative adversarial networks (GAN) to create load profiles [11]. For instance, Song et al. proposed ProfileSRGAN, which aims to up-sample load profiles from low-resolution to high-resolution, thereby restoring high-frequency components lost during the down-sampling process [12]. However, the physics of the system were not considered in their formulation. If we overlook the inherent dynamics of the power system (such as the temporal correlations), the generated dataset may exhibit significant fluctuations for a given time frame, making them unrealistic. Thus, embedding the system's knowledge into the learning model by incorporating appropriate constraints during training of the generative models is crucial.

In the context of deep learning, constrained optimization is receiving significant attention as it is being widely recognized that simple constraints can be effectively incorporated within deep learning models to improve performance [13]. Constrained optimization can be done by adding penalty terms to the loss function of the deep learning model in order to enforce soft boundary constraints [14]. However, such approaches only penalize infeasibility without guaranteeing feasibility. Alternatively, hard constraints can be enforced by either processing the output of an unconstrained model [15], or designing a model that inherently produces feasible predictions [16]. Most recent works have investigated differentiable optimization layers for neural networks (NNs), as such approaches could be used to directly enforce the constraints, e.g., by projecting NN outputs onto a constraint set using quadratic programming layers in the case of linear constraints [17], or convex optimization layers in the case of convex constraints [18].

### A. Summary of Our Contributions

In this study, we introduce an innovative approach using GAN to create high-resolution temporally correlated power consumption datasets that mimic actual load injections. This method is particularly beneficial when the available training data lacks detail or frequency. Our main contributions are:

1. **Incorporating constraints within the GAN**: By using a convex optimization (CVXPY) layer, we embed a set of rules within our GAN that dictate how its internal components should interact. This ensures that the data produced by the GAN adheres closely to realistic patterns and constraints, enhancing the quality and applicability of the generated datasets.
2. **Leveraging slow timescale data for high-resolution output:** Our model produces high-resolution (1-minute interval) load profiles using average power consumption data obtained at 15-minute intervals from smart meters. This addresses the challenges posed by the scarcity of high-resolution training datasets in distribution systems.
3. **Exhibiting temporal consistency:** The generated profiles not only exhibit statistical realism but also maintain temporal consistency. This makes them suitable for important applications such as high-speed DSSE and fast voltage control.

## II. BASIC GAN MODEL

GAN is made up of two NNs that compete against each other. The first NN is the *generator*, $G$, which creates synthetic samples, while the second NN is the *discriminator* (or critic), $D$, that distinguishes between the real and generated samples [19]. The primary goal of GAN is to generate new statistically similar (but not identical) samples for an existing (real) dataset by first learning the distribution of the real dataset, and then mapping it to a separate latent space. The initial focus is on optimizing the $D$, given the $G$. The training process for the $D$ entails minimizing the cross-entropy loss, which is formulated as shown below [20]:

$$\text{Loss}(D) = -\frac{1}{2} E_{r \sim p_{\text{data}}(r)}[\log D(r)] - \frac{1}{2} E_{f \sim p_f(f)}\left[\log\left(1 - D(G(f))\right)\right] \quad (1)$$

where, $r$ is sampled from real data with probability $p_{\text{data}}(r)$, $f$ is sampled from the prior distribution $p_f(f)$ such as uniform or Gaussian, and $E$ denotes the expectation operation. The training data consists of two parts: one obtained from the real data distribution $p_{\text{data}}(r)$, while the other obtained from the generated data distribution $p_G(r)$. Given the $G$, we minimize (1) to obtain the optimal solution. Now, (1) can be reformulated as shown below:

$$\text{Loss}(D) = -\frac{1}{2} \int_r p_{\text{data}}(r) \log D(r) \, dr - \frac{1}{2} \int_f p_f(f) \log\left(1 - D(G(f))\right) df$$
$$= -\frac{1}{2} \int_r [p_{\text{data}}(r) \log D(r) + p_G(r) \log(1 - D(r))] dr \quad (2)$$

where, $D(r)$ denotes the probability of $r$ being sampled from the real data rather than the generated data. Now, for any $(a, b) \in \mathbb{R}^2 \setminus \{0,0\}$ and $c \in [0,1]$, the expression: $-a \log(c) - b \log(1 - c)$ achieves its minimum value at $c = a/(a + b)$. Hence, given $G$, (2) achieves its minimum value at:

$$D^*_{G(r)} = \frac{p_{\text{data}}(r)}{p_{\text{data}}(r) + p_G(r)} \quad (3)$$

From the optimal solution shown in (3), it can be realized that the $D$ of the GAN estimates the ratio of two probability densities. When the input data is from the real data $r$, the $D$ strives to make $D(r)$ approach one. Conversely, if the input data is from the generated data $G(f)$, the $D$ strives to make $D(G(f))$ approach zero, while the $G$ tries to make it approach one. Since this is a zero-sum game between $G$ and $D$, the loss function of $G$ is the negative of the loss function of $D$; i.e., $\text{Loss}(G) = -\text{Loss}(D)$. Therefore, the overall optimization formulation of the GAN can be expressed as a two-player minimax game with value function $V(G, D)$, as shown below:

$$\min_G \max_D V(G, D) =$$
$$E_{r \sim p_{\text{data}}(r)}[\log(D(r))] - E_{\sim p_f(f)}[\log(1 - D(G(f)))] \quad (4)$$

Finally, we must train $G$ to minimize $\log(1 - D(G(f)))$. To do this, we fix $G$ and optimize $D$ to maximize the discrimination accuracy of $D$. Then, we fix $D$ and optimize $G$ to minimize the discrimination accuracy of $D$, as seen in (4). This process alternates, with the global optimal solution achieved only when $p_{\text{data}} = p_G$. Note that during the training process, the parameters of $D$ undergo empirical updates multiple times before the parameters of $G$ are updated. This sequential updating strategy contributes to the convergence of the GAN and the attainment of a more effective generative model. For a detailed explanation of the working of GAN, please see [19].

### III. Proposed Constrained GAN

Without any loss of generality, it can be assumed that the slow timescale aggregated AMI data are $m$ times slower than the fast timescale data, say, coming from SCADA. Since the SCADA system is not deployed throughout the distribution system to observe it at high-speeds, the goal is to *create realistic statistically accurate SCADA-like data using smart meter measurements*. In the proposed approach, we use historical average power data from smart meters to train a GAN with inter-neuron constraints inside the generator network through the CVXPY layer. The generator produces fast timescale synthetic profiles of size $m \times s$ from the CVXPY layer, which are then aggregated column-wise to generate $s$ smart meter-like measurements for comparison by the discriminator.

#### A. Generator Block of Proposed GAN

The proposed GAN is trained by incorporating a CVXPY layer after the dense layers of the generator's deep NN (see Fig. 1). This creates an end-to-end trainable network that enables addition of inter-neuron constraints. The CVXPY layer is a convex optimization model that trains the GAN with user-defined inputs. Note that the CVXPY layer is the penultimate layer of the generator. After this layer, we have the *Aggregator function* layer, which is non-trainable. This layer performs the average operation on the outputs of the CVXPY layer to match the generator's outputs with the dimensions of the slow timescale AMI data. Finally, the aggregated data is sent to the discriminator of the proposed GAN. The input of the generator is a noise vector that is sampled from a uniform distribution.

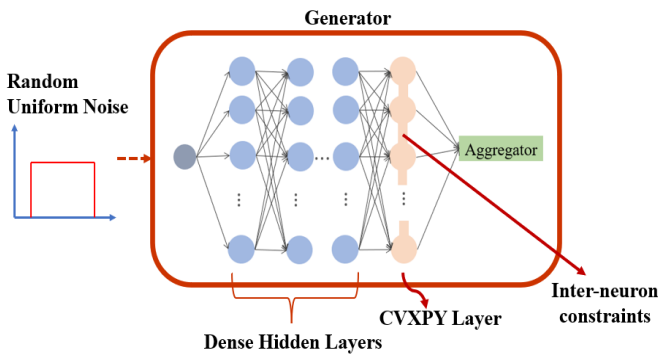

Figure 1. Proposed generator network model.

#### B. CVXPY Layer

The CVXPY layer is a differentiable optimization layer [18]. Optimization layers add domain-specific knowledge or learnable hard constraints to machine learning (ML) models. They solve convex, constrained optimization problems of the form:

$$x(\theta) = \arg\min_x J(x, \theta) \text{ such that}$$
$$l(x; \theta) \leq 0$$
$$h(x; \theta) = 0 \quad (5)$$

with objective $J$ and constraint functions $l$ and $h$ doing end-to-end learning with respect to parameter $\theta$. These constraints are vital in scenarios where the desired outcomes of a model must satisfy particular conditions.

In our case, the CVXPY layer optimizes the generator's NN as it ensures creation of synthetic data that satisfies conditions specified via the inter-neuron constraints. The imposition of such constraints elevates the generator's capability from merely mimicking data patterns to actively conforming to the intricate dynamics of the system. The proposed CVXPY layer solves a parameterized convex problem in the forward pass to produce a solution. In the backward pass, it computes the derivative of the solution with respect to the parameters. Through this process, the CVXPY layer learns a parameterized objective function and multiple hard constraints from data that are initially unknown to the model. The details of the objective function and constraints is provided in Section III.D.

#### C. Discriminator Block of Proposed GAN

The discriminator is a vanilla deep NN with dense layers and a Sigmoid activation function as its output. It acts similar to a switch (see Fig. 2). The discriminator is first trained with actual AMI data. Subsequently, it is trained with the fake dataset created from the generator. At the end, it outputs a continuous probability score indicating the likelihood that a given input is real or fake.

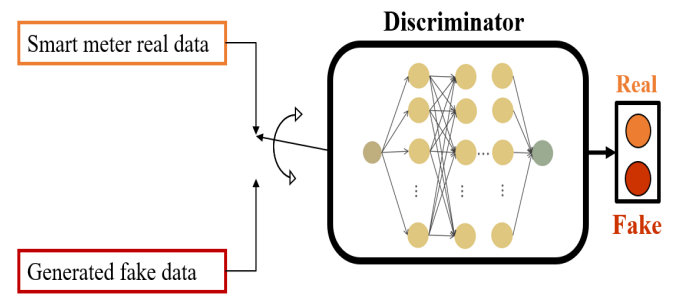

Figure 2. Discriminator network model.

#### D. Overall Structure of the Proposed GAN

The objective function of the proposed GAN model is a two-player minimax game given by:

$$\min_G \max_D V(G, D) = E_x[\log(D(x))] - E_z[\log(1 - D(G(\sum \text{CVXPY}(z))))] \quad (6)$$

Equation (6) is obtained by modifying the objective function of the basic GAN model (see (4)) by including the CVXPY and Aggregator function layers inside the GAN framework. The final structure is shown in Fig. 3. Both generator and discriminator have two hidden (dense) layers with 128 neurons in each layer. Batch normalization and dropout rate of 0.3 are used in these layers. The third hidden layer of the generator is the CVXPY layer, which contains 15 neurons and is followed by an Aggregator as the output layer, which has non-trainable parameters. Note that this structure corresponds to a SCADA availability of one sample every minute, while AMI data is assumed to be available for 15 aggregated samples, i.e., $m = 15$.

To learn the fast timescale load distribution using the smart meter measurements, we made the following modifications. As an alternative to generating *one* sample from the CVXPY layer, we generated 15 samples from it (since $m = 15$) and averaged them to imitate a measurement from smart meters at each slow timescale interval. Next, to embed temporal correlations into our data generation process, the following constraints were added in the CVXPY layer:

$$z^* = \underset{z}{\mathrm{argmin}} \sum_{i=1}^{m}((z_i - a_i)_i)^2 \quad (7.1)$$

$$k_1 . L \leq z_i \leq k_2 U \quad (7.2)$$

$$(1 - k_3) z_{i-1} \leq z_i \leq (1 + k_3) z_{i-1} \quad (7.3)$$

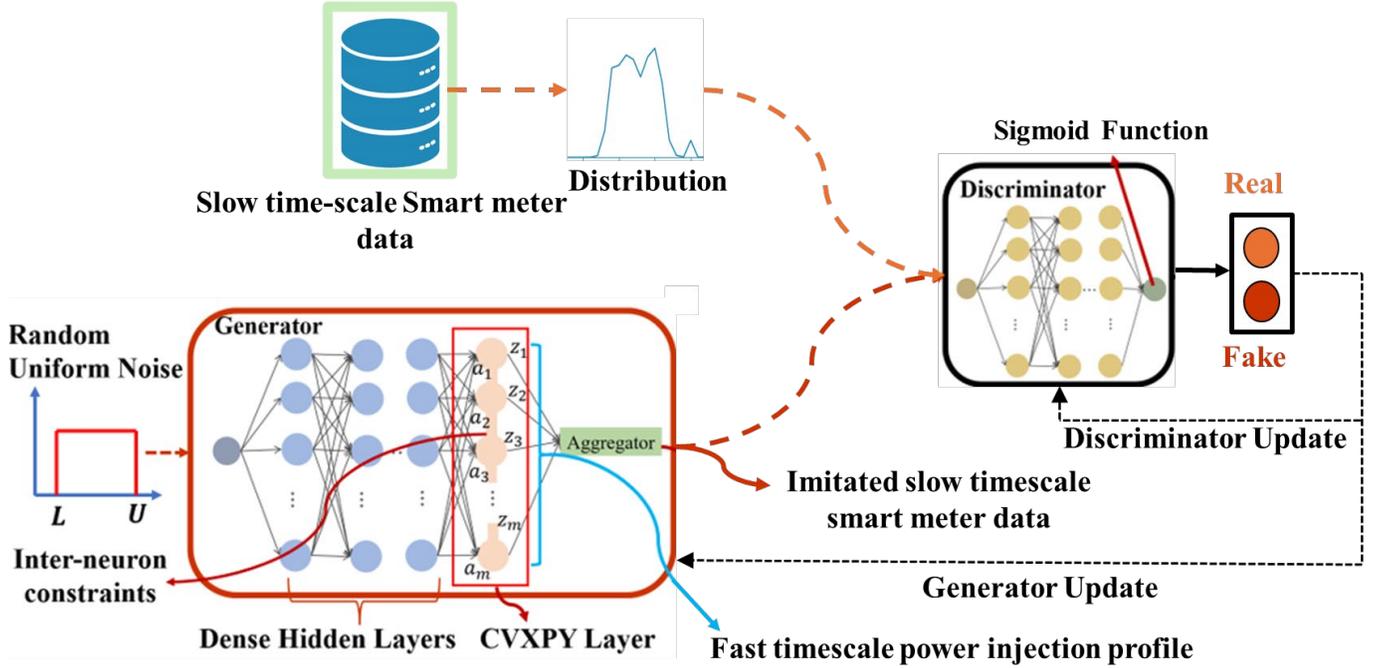

Figure 3. Proposed constrained GAN model.

Equation (7.1) represents the core objective of minimizing the sum of squared deviations between the generated samples $z_i$ and the inputs $a_i$, where $z^*$ denotes the optimized output vector, and $a_i$ represents the input values to the CVXPY layer. This process aims to generate power consumption values that are closely aligned with the inputs. Next, to ensure the generated values remain within practical bounds, the model imposes two critical constraints. First, in (7.2), each $z_i$ is forced to lie between a lower bound and an upper bound, scaled from historical minimum ($L$) and maximum ($U$) values by factors $k_1$ and $k_2$, respectively. This scaling accommodates expected fluctuations in power usage while maintaining adherence to observed historical data. Secondly, (7.3) imposes restrictions on the rate of change from one sample to the next. It ensures that each generated sample $z_i$ does not deviate from its predecessor $z_{i-1}$ by more than a predetermined percentage, encapsulated by the parameter $k_3$. This constraint is pivotal for capturing the inherent energy consumption temporal patterns, where drastic shifts within small time intervals do not occur. The parameters $k_1$, $k_2$, and $k_3$, along with the bounds $L$ and $U$, are user-defined quantities. For obtaining practical insights into the values that can be used for these parameters, please see the next section.

## IV. SIMULATION RESULTS

We used the profiles of load injection from the Pecan Street (PS) database [21] for describing the operation of our GAN. The load-level measurements occurring at 1-minute and 15-minute intervals from this database were employed in this analysis. Note that only the 15-minute data was used to train the GAN, while the 1-minute data was used to determine the values of the parameters $k_1$, $k_2$, $k_3$, $L$, and $U$. Since PS data represents actual data from a distribution system, the values of the parameters calculated from the PS data can be treated as a representative of other distribution systems as well in which such fast timescale data are not available.

Next, to further increase efficacy of the deep learning model, samples were selected from historical data with similar features, such as same season and hour of the day. Therefore, for each load of the PS database, we selected load data from June 1st through August 31st of 2018 between 12 Noon and 1 PM. This targeted approach allowed us to train the GAN with highly relevant samples. After training using the 15-minute interval data, we employed the GAN's generator to create multiple high-resolution, temporally-correlated load profiles. To validate the generated profiles, we compared their cumulative distribution function (CDF) with that of the 15-minute smart meter readings from the PS database. Fig. 4 shows the CDFs for a particular load. The comparison revealed a close match between the CDFs, indicating the generator's ability to effectively replicate load injections.

Note that although the generator produces outputs at the slow timescale rate of one sample per 15 minutes (i.e., the output of the Aggregator layer in Fig. 3), the fast timescale measurements at 1-minute intervals, namely $z$, are available at the output of the CVXPY layer (penultimate layer). This output *is* the required SCADA-like data that the proposed

GAN can produce at every location where a smart meter is placed. Thus, our method ensures SCADA-timescale observability for all loads with smart meters (irrespective of the load model or load type).

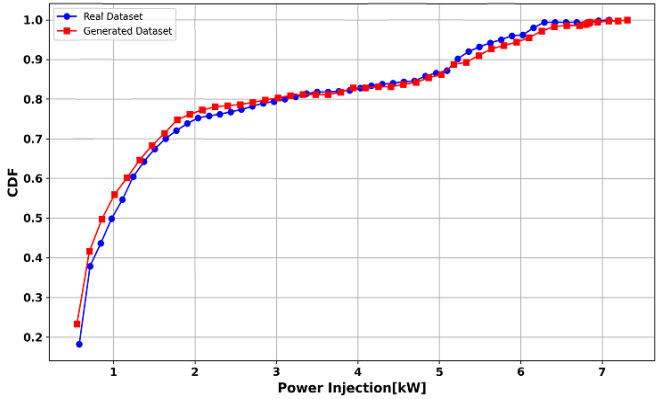

Figure 4. CDF comparison. Blue curve is the CDF of 15-minute data; Red curve is the CDF of imitated 15-minute measurements from learnt distribution.

Next, we compare the proposed *constrained* GAN with the traditional GAN developed in [5]. This comparison is meant to highlight the significance of the CVXPY layer in enforcing essential inter-neuron constraints that play a pivotal role in the *fidelity* of the generated data. The method proposed in [5] demonstrated high degree of similarity in the *aggregated values*, reaffirming the overall effectiveness of using GAN in capturing statistical characteristics of AMI data (similar to what was shown in Fig. 4). However, a significant difference emerges when we delve into the fast timescale measurements, as the traditional GAN framework is unaware of the power system constraints. This is elaborated below.

For the same load (that was used for Fig. 4), it was observed that for the season and hour that were the focus of this study, the 1-minute power injection data varied between 0 and 10.26 kW, while the aggregated 15-minute AMI data varied between 0.55 and 7.38 kW. Therefore, we chose $k_1 = 0\%$, $k_2 = 139\%$, $L = 0.55$, and $U = 7.38$. Furthermore, it was observed that a valid 1-minute measurement never changed by more than 50% of its predecessor; as such, we put $k_3 = 50\%$. Note that the choice of the values for these parameters depends on the selected load and would be different if another load had been chosen instead. The proposed constrained GAN was able to account for the power system constraints through the CVXPY layer, while the traditional GAN (that did not have the CVXPY layer) could not. The difference is observed in Fig. 5, which shows 1-minute samples for a period of 15 minutes.

As the traditional GAN is not aware of the constraints present in the load injection profiles, it often outputs values that are outrageous (e.g., 385% difference between two consecutive generated samples). However, the proposed GAN generates much more reasonable (realistic) values by enforcing the physical constraints during high resolution sample generation. Moreover, it is worth noting that the averages of both profiles are relatively close to each other, with the constrained GAN profile averaging 2.48 kW and the traditional GAN averaging 2.9 kW. This highlights that although 15-minute data can show similar values for both the profiles, their corresponding high-resolution profiles can be significantly different, leading to severe violations of realistic load consumption profiles. Therefore, while matching CDFs for 15-minute intervals is necessary (as shown in Fig. 4), additional changes must be made into the structure of the GAN (such as the inclusion of the proposed constraints) to ensure that the high-resolution 1-minute data produced by them are realistic.

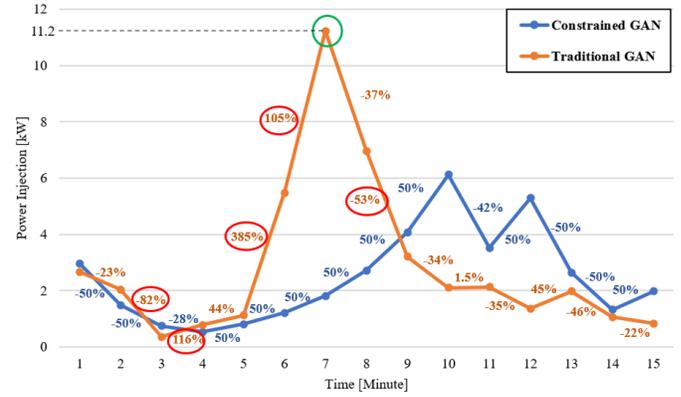

Figure 5. A high-resolution instantaneous power injection profile with traditional GAN and constrained GAN. Numbers on each line segment show the percent change for two consecutive samples. Green circle shows violation of (7.2) and red circles show violation of (7.3) for the traditional GAN.

To further evaluate the performance of our method in comparison to the traditional GAN, we analyzed a dataset consisting of 500 synthetic load profiles, each comprising 1-minute sequence of 15-minute interval values. This allowed us to calculate the percentage increase or decrease between two consecutive minutes over a statistically large dataset. The key differences between the two GANs are readily apparent in Table I. For the proposed constrained GAN, the maximum percentage increase or decrease for any two consecutive minutes within each 15-minute time-frame consistently remained at or below 50% (see second column). This finding attests to the efficacy of our approach in enforcing the constraint between consecutive time intervals via the CVXPY layer. Conversely, for the traditional GAN, the percentage increase or decrease for two consecutive time-frames could be extremely large (e.g., 5000%) resulting in abrupt changes in the generated consecutive power injection values, which is not realistic. In addition, the third column in Table I indicates that for the traditional GAN, we do not have any control on the maximum value generated for high resolution power injections (it went up to 13.21 kW in our simulations), while by adding the CVXPY layer we were able to limit the maximum value for generated high-resolution power injections to 10.26 kW (which matches with the limits of the actual 1-minute data as seen in the third row). Lastly, it is crucial to highlight that the ground truth values in Table I come from the PS database, where observed maximum increases and decreases in load data, as well as the peak load injection values, closely match the performance of our proposed constrained GAN. This parity underscores the enhanced realism and practical applicability of our model, aligning closely with empirical data patterns and supporting the need for incorporating physical constraints into synthetic data generation frameworks.

The proposed constrained GAN had an overall complexity of $O(n^3)$, where $n$ denotes the numbers of variables involved. This complexity, combined with the iterative nature of GAN training, contributes to a training time of approximately 6 hours on a computer with 256 GB RAM, Intel Xeon 6246R CPU @3.40GHz, Nvidia Quadro RTX 5000 16 GB GPU. However, this is not a major concern since the selected application is an offline problem.

Table I. Maximum percentage change for two consecutive samples and maximum instantaneous value for 500 high-resolution power injection profiles

| 1-minute resolution data | Maximum decrease/increase [%] | Maximum power injection [kW] |
|---|---|---|
| Traditional GAN | 5000 | 13.21 |
| Constrained GAN | 50 | 10.26 |
| Ground Truth | 50 | 10.26 |

## V. Conclusion

In this work, we proposed a novel GAN-based method for generating, from AMI measurements, SCADA-like fast time-scale data that currently do not exist in the secondary side of the distribution system. A key feature of the proposed GAN is that it learns a distribution when the samples are not directly observable. Thus, the generated data can be used as training data for ML-based DSSE and voltage control algorithms.

By embedding convex optimization layers into the proposed GAN framework, the proposed approach is able to create datasets that conform to specific constraints that were previously impossible to achieve. This new methodology holds great promise for various applications that require high-quality datasets with specific domain knowledge embedded inside the training process. It also opens up new possibilities for the development of more robust and accurate ML models that can manage complex power system tasks. The future work will involve focusing on generating higher-resolution instantaneous power injection profiles with both temporal and spatial correlations and using them for enhanced monitoring and control of active distribution systems.

## Disclaimer

The author(s) is/are solely responsible for the content of this article. Publication of this article shall not constitute or be deemed to constitute that Eversource Energy has reviewed, approved, or endorsed the article content or that the data presented therein are correct or sufficient to support the conclusions reached or that the experiment design or methodology is adequate.